\documentclass[aps,prl,groupedaddress,showpacs,twocolumn,floatfix]{revtex4}
\usepackage{amsmath}
\usepackage{amssymb}
\usepackage{graphicx}
\usepackage{epsfig}

\begin{document}

\title{Excitation Spectra of Correlated Lattice Bosons in a Confining 
Trap} 
\author{Dmitry L. Kovrizhin $^{1}$,  G. Venketeswara Pai $^{2}$, and 
Subhasis Sinha $^{3}$}
\address{$^{1}$ Clarendon Laboratory, Parks Road,
University of Oxford,
Oxford OX1 3PU, United Kingdom \\
$^{2}$Department of Physics, Technion - Israel Institute of
Technology, 32000 Haifa, Israel \\
$^{3}$ S. N. Bose National Centre for Basic Sciences, Block JD, Sector III, 
Kolkata 700098, India }

\begin{abstract}

We consider ultracold bosonic atoms in an optical lattice with
an external trapping potential. To study the excitation energies of the
resulting Bose-Hubbard model, we develop a method based on a
time-dependent generalization of the Gutzwiller ansatz.
We calculate the excitations of the homogeneous system both in
insulating and superfluid regime, concentrating particularly on those
near the superfluid-Mott insulator boundary. Low-lying excitation energies
in presence of a static harmonic trap are obtained using this method
and compared with the homogeneous case. Further, we explore the dynamics
of the center of mass and the breathing mode in response to
time-dependent perturbations of the trap.

\end{abstract}


\maketitle
\narrowtext
The study of ultracold bosonic atoms confined in an optical lattice
as a correlated system \cite{jaksch} has been an active area of research in the
past few years \cite{review}. It could provide a better
understanding of many phenomena driven
by strong many-body correlations in a controlled fashion, which is
rather difficult to explore in conventional solid state systems.
These include the quantum phase transition between the Mott insulating (MI)
and the superfluid (SF) phases \cite{greiner, esslinger1}, Luttinger liquid
behavior in quasi-one dimensional condensates
in the Tonks-Girardeau regime \cite{TGnature}, etc. It has 
been proposed that
these systems could be used to simulate various quantum spin models
\cite{spin}, especially in reduced dimensionality and for creation 
of frustrated lattices
to explore the possibility of unconventional quantum phases
\cite{senthil}. It may also be possible to achieve reliable prototypes
for quantum computing and information processing \cite{qcomp}.
Recent experiments involving dipolar atoms \cite{Cr} are expected to
lead to the realization of new phases of matter such as the supersolid (SS)
\cite{goral, dima}, whose existence is still not unambiguous
in solid $^4$He \cite{chan}.

The MI-SF transition in lattice bosons is controlled by the competition
between the interaction among them and their kinetic
energy \cite{fisher}.
However, the translational symmetry breaking confining potential,
that is present in experiments, would lead to inhomogeneous phases and
even phase coexistence for appropriate parameters.
There have been various proposals to observe such
structures \cite{coexistence}. A clear indication of
it will be their excitation spectra, which will be
different from their homogeneous counterparts and the way the system
behaves under time-dependent perturbations of the trap \cite{clark}. 
Further, the study of the collective modes has become an important
experimental tool to analyze correlation properties of ultracold 
quantum gases in a trap \cite{esslinger2,esslinger1}.
With this
in mind, we systematically explore the excitations of
parabolically trapped lattice bosons from deep SF phase to insulating phase. 
Our method also provides
a detailed microscopic understanding of the behavior of the trapped SF under
perturbations which, hitherto, has been studied using the 
Gross-Pitaevskii equation (GPE) or equivalent hydrodynamic approach
\cite{stringari}. These methods are valid only when the SF fraction is
very large; they fail to capture the modified physics both in the homogeneous
SF near the MI boundary and in the inhomogeneous phase coexistence regime.

The harmonically trapped
alkali atoms with short range interactions in an optical lattice can be
modeled by the effective Bose-Hubbard model
(BHM) \cite{fisher,jaksch}:
\begin{equation}
\hat{H}=-t\sum_{i,\delta}\hat{a}_{i}^{\dag}\hat{a}_{i+\delta}+\sum_{i}\left[
\frac{U}
{2} \hat{n}_{i}\left(  \hat{n}_{i}-1\right)
+{{m \omega^2} \over 2} {\bf r}^2_i
\hat{n}_{i}\right]. \label{1}
\end{equation}
The three terms represent the lattice kinetic energy, the on-site
interaction, and the harmonic trap potential respectively.
The operator $\hat{a}_{i}^{\dag}$ creates a boson at site $i$,
$\hat{n}_{i}=\hat{a}_i^{\dag}\hat{a}_i$ is the boson number operator,
${\bf r}_i$ is the distance of the site $i$ from the minimum of the
trap potential with frequency $\omega$;
$t, U$ are the nearest-neighbor hopping and the on-site
Hubbard interaction respectively, $m$ is the mass of the atoms, and  $\delta$ 
represents the nearest
neighbors of the site $i$. This Hamiltonian without any confining
potential has been studied extensively using a variety of methods
such as the mean field theory \cite{fisher}, quantum
Monte-Carlo \cite{qmc}, perturbation series \cite{monien} and density matrix
renormalization group (DMRG) \cite{dmrg,tdmrg}. In the absence
of trap, the ground state is an incompressible MI with an integer
number of atoms at every lattice site when the interactions
dominate and the particle density is commensurate with the lattice.
When the kinetic energy dominates, and in general when the density is
incommensurate, a SF ground state is obtained.
The low-lying excitations in the MI state are gapped particle-hole
excitations. The SF phase has gapless,
acoustic, Bogoliubov quasiparticles. For large enough $t/U$, the excitations
can be described by the discrete non-linear Schr\"{o}dinger(DNLS) 
equation (the lattice version of the GPE)
\cite{stringari}, which becomes progressively inaccurate
as one moves towards the MI-SF boundary \cite{dima}.
In presence of a trap 
one encounters an inhomogeneous SF phase and at smaller values of $t/U$,
the renormalized chemical potential gives rise to alternating shells
of SF and MI regions. 
In absence of {\it 
numerically exact} calculations, which are not hindered by limited system size
or small number of bosons, a (linearized) time 
dependent variational mean field analysis
is an appropriate tool to explore the excitations of these systems; it gives
quantitatively correct results \cite{qmc} in homogeneous case.
Hence, we use the variational Gutzwiller mean field approach to study the
ground state and use a modified time-dependent Gutzwiller ansatz 
to study the low-lying excitation energies. In particular, we concentrate on
the excitations in the homogeneous phase near the MI-SF boundary, the
low-lying spectra in the presence of a static trap as a function of
parameters, and the dynamics of the system in response to time-dependent
perturbations of trap potential.

We calculate the inhomogeneous ground states of the above
model, in general in $d$ dimensions, for a given set of parameters and 
the chemical potential $\mu$ by minimizing
$\left < \Psi \left| \hat{H} - \mu \hat{N} \right| \Psi \right >$ 
with a Gutzwiller wavefunction
$\left| \Psi \right> = \prod_{i}\sum_n f_n^{(i)} \left| n,i \right>$ 
with respect to the variational parameters $f_n^{(i)}$, where
$\left| n,i \right>$ is the Fock state with $n$ particles at site $i$
and $\hat{N} = \sum_i \hat{n}_i$ is the total particle number operator.
The excitation energies above the ground state are, then, obtained from the
real-space dynamical Gutzwiller approach with the variational parameters
$f_n^{(i)}$ being time dependent that was introduced 
in Ref. \cite{timedependent}
and modified for the calculation of excitation spectra in Ref. \cite{dima}.
Minimization of the effective action
$\left <\Psi \left|i\hbar {\partial \over {\partial \tau}}-
\hat{H} +\mu\hat{N}\right|\Psi \right >$ gives the equations of motion
for $f_n^{(i)}$:
\begin{eqnarray}
i\hbar \frac{\partial {f}_{n}^{\left(  i\right)  }}{\partial \tau} = 
\left[  
\frac{U}{2}n_i\left(  
n_i-1\right)
-\mu n_i + {{m \omega^2} \over 2} {\bf r}^2_i n_i \right]  
f_{n}^{\left(  i\right)}\nonumber\\
-t \sum_{\delta}\left[\phi_{i + \delta}^{\ast}\sqrt{n+1}f_{n+1}^{\left(  
i\right)}+\phi_{i + \delta}
\sqrt{n}f_{n-1}^{\left(  i\right)  }\right]. \label{2}
\end{eqnarray}
Here $\phi_i = \left< \Psi \left| \hat{a}_{i}
\right| \Psi \right>$ is the local condensate (SF) order parameter and $\tau$
is time. The oscillation frequencies of the small amplitude fluctuations 
$\delta f_n^{(i)}(\tau)$ around
the ground state $\bar{f}_n^{(i)}$ give the excitation spectrum.
Normalization of the wavefunctions, $\sum_{n} |f_n^{(i)}(\tau)|^2 = 1$,
at each site is enforced using a Lagrange multiplier $\lambda_i$.

Firstly, we consider the excitations in the homogeneous system without a 
trap.
For the MI phase with $n_0$ bosons per site, the ground state is defined by 
$\bar{f}_n^{(i)} =
\delta_{n,n_0}$. The resulting linearized equations
lead to particle(p) and hole(h) excitations with
dispersions
\begin{equation}
\varepsilon_{p(h)} = 
\sqrt{\frac{U^2}{4} + \frac{\epsilon_{\bf k}^2}{4} + \epsilon_{\bf k} U (n_0 + \frac{1}{2} )}
\pm \left[U(n_0 -\frac{1}{2}) -\mu +\frac{\epsilon_{\bf k}}{2}\right],
\end{equation}
where $\pm$ corresponds to p(h) excitations and
$\epsilon_{\bf k} =  - 2 t \sum_{j = 1}^{d} {\mathrm{cos}}(k_j)$. This is in 
agreement with the results obtained from the slave-boson mean-field 
theory \cite{stoof} and random-phase approximation \cite{sengupta}.
As one approaches the phase boundary from the MI side,
the energy gap for p(h) excitations gradually decreases, vanishing at the
transition.
The phase boundary at which the MI-SF transition occurs is obtained from the 
condition of vanishing energy-gap,
$\varepsilon_{p(h)}({\bf k=0}) = 0$, and is in agreement with the calculations
using second 
order perturbation theory in $t/U$ \cite{monien}. 
Deep in the SF
phase (i.e., when $t \gg U$), and for large SF fraction
(${\left|\phi_i\right|}^2 \sim n_{i}$), the wavefunction at each site
can be represented by coherent states, i.e.,
$f_n^{(i)} = \phi_i^n e^{-{\left|\phi_i\right|}^2/2}/\sqrt{n!}$ and 
Eq.(2) reduces to a DNLS equation for the classical field $\phi_i$:
\begin{eqnarray}
i \hbar {{\partial \phi_i} \over {\partial \tau}} &=& -t \sum_\delta \phi_{i+\delta}
+ U {\left| \phi_i \right|}^2 \phi_i
- \mu \phi_i + {{m \omega^2} \over 2}
{\bf r}_i^2 \phi_i. 
\label{3}
\end{eqnarray}
In the absence of a trap, this gives gapless acoustic mode with sound velocity
$c_s = |\phi|a\sqrt{2t U}/\hbar$, where $a$ is the lattice spacing. 
The MI-SF transition near the
Mott lobe at a commensurate density $n_0$ can be understood {\it via}
a simple three-state variational ansatz at site $i$:
$\left|\psi_i\right> = f_{1}\left|n_0-1\right> + f_{0}\left|n_0\right> + 
f_{2}\left|n_0 + 1\right>$.
This captures the build-up of number fluctuations \cite{huber},
and hence the phase coherence as the system makes
a transition to the SF phase at a critical coupling $(U/2td)_c = {(\sqrt{n_0} + 
\sqrt{n_0+1})}^2$. Introducing time-dependent fluctuations
in the variational parameters lead to the Bogoliubov spectrum in the SF phase 
with the sound velocity $c_s$ given by,
\begin{equation} 
c_s = t\sqrt{d} \cos \theta \sqrt{(\alpha^2 \cos^2 \theta 
-1)/2},
\end{equation}
where $\alpha = {(\sqrt{n_0} + \sqrt{n_0+1})}^2 = U/(2td \cos 2\theta)$.
Interestingly, at the boundary where $\theta=0$, the sound velocity
is finite, although the SF order parameter $\phi$ vanishes. This was first 
pointed out in Ref. \cite{altman} with $n_0\gg 1$ and 
an effective relativistic GPE has been proposed to 
capture the dynamics close to the MI phase. For small momenta,
the (gapped) amplitude mode \cite{altman} in the SF has the dispersion
$\omega({\bf k})=\sqrt{\Delta^2+c^2{\bf k}^2}$ with
$\Delta = td \sin 2\theta \sqrt{\alpha^2-1}$ and
$c^2 = (t^2d/8) \left[ 9+ \left(4 \alpha^2-13\right) \cos^2 2 \theta
\right]$ in $d$-dimensions. At the boundary, this mode becomes degenerate
with the gapless, Bogoliubov mode.
However, deep in the SF phase three state variational ansatz is insufficient 
and we increase the number of Fock states at each site and 
numerically solve 
the linearized equations for the fluctuations
$\delta f^{(i)}_{n}$ to obtain the collective mode with momentum $k$.
The dispersion of the lowest acoustic mode in the SF phase is shown in Fig.1 
for different coupling strengths.
We compare our result with those obtained from DNLS and find good 
agreement deep in the SF regime ($|\phi_i|^2 \sim n_i$),
whereas DNLS fails near the MI boundary (see Fig.1).

Main advantage of our method is that it is simple and efficient enough to
implement in confined geometries and for inhomogeneous systems. 
Confining traps which are present in experiments
can be approximated by a harmonic oscillator potential
$V({\bf r}_{i})=\frac{1}{2} m \omega^2 {\bf r}_{i}^2$.
Minimization of free energy gives, in general, inhomogeneous ground state
that show coexisting SF and MI phases (with
step-like structure) for certain parameters. Compressible edge forms on the
boundary of such droplet.
Having found the ground state
for a given $\mu$, we calculate the low-lying
spectrum by solving the system of linearized equations (Eq.(2)).
Recently collective excitations of BHM in a 1D harmonic trap have been 
calculated 
by exact diagonalization of the Hamiltonian for small system size \cite{lundh}.
However, the complexity of the problem grows exponentially with
increasing number of the lattice sites and boson filling fractions.
Numerically efficient T-DMRG methods \cite{tdmrg,clark} are restricted to
1D systems with short range interactions, whereas our method
can be applied to higher dimensions, where the mean field method is 
expected to work better.
\begin{figure}[htb]
\epsfig{file=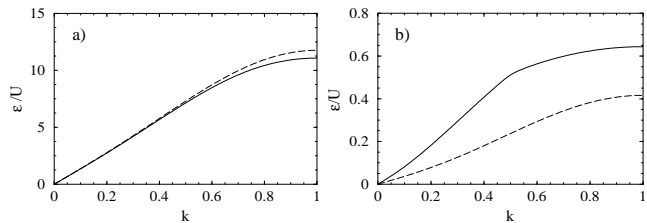,width=8.5cm}
\vspace*{-0.5cm}
\caption{
Excitation spectrum of the uniform BHM with $\mu/U = 0.5$, a) for $t/U=2$ and 
b) for $t/U=0.087$. For comparison dotted lines represent excitation
spectrum obtained from DNLS. Momentum $k$ is in units of $\pi/a$.
}
\label{fig:1}
\end{figure}
\begin{figure}[htb]
\epsfig{file=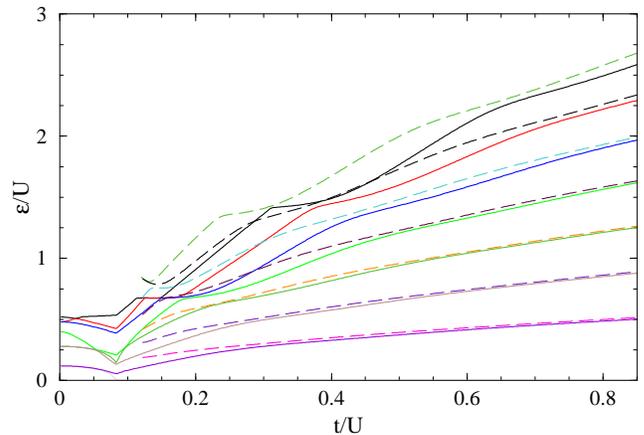,width=8.5cm}
\vspace*{-0.5cm}
\caption{Few low-lying excitations of the BHM in a trap with $m\omega^2a^2/U=0.16$
and $\mu/U=0.6$ (solid lines). For comparison the dotted lines denote
the energies obtained from DNLS for the same parameters.}
\label{fig:2}
\end{figure}
In Fig.2 we show a few low-lying excitations of bosons
in a 1D harmonic trap, as a function of $t/U$ for a fixed $\mu/U$.
We clearly notice three different regimes by changing the interaction
strength $t/U$. In the deep SF region (i.e., $t/U\gg 1$), excitations are
Bogoliubov quasiparticles and their energies asymptotically match with those
obtained from DNLS. In this regime 
the hopping parameter $t$ is equal to the kinetic energy
$\hbar^2/(2 m a^2)$.
For large $t/U$, the $n$th excitation branch approaches 
$\omega\sqrt{n (n + 1) ta^2 m}$ asymptotically 
and is in agreement with the hydrodynamic modes
of a dilute Bose gas in 1D harmonic trap without a lattice \cite{menotti}.  
As the system enters the correlated regime (for smaller $t/U$), 
which is close to the phase boundary,
the excitation energies start to deviate from those obtained from DNLS 
and
many avoided level crossings occur in the energy spectrum.
We notice the clear signature of particle and hole type excitations 
deep in the MI phase ($t/U \ll 1$).
In contrast to the homogeneous BHM, where p(h) energies are
site-degenerate, in case of harmonically trapped BHM, they split
and become site-dependent in the insulating phase.
It is easy to understand these p(h) excitation energies in the atomic
limit of BHM.
For $t=0$ and $\mu < U$ the system forms a droplet of size $2 n_{max}$ with 
all sites $i$ for $-n_{max}\leq i \leq n_{max}$
filled with one particle per site, where $n_{max}$ being the integer 
part of $\sqrt{2 \mu/(m\omega^2 a^2)}$.
Thus in presence of the trap, the energy for adding or removing a
particle, in general, depends on the site index $i$, and is given by
$E_p = U - \mu + \frac{1}{2}m\omega^2 a^2 i^2$ and
$E_h = \mu - \frac{1}{2}m\omega^2 a^2 i^2$.
At the edge of the trap (at $i= \pm n_{max}$),
$E_p \approx U$ and $E_h \approx 0$.
Similarly,
it costs very little energy to create a particle at the empty site just
outside the droplet (i.e., $i = \pm (n_{max} + 1)$). 
This leads to the formation of the gapless edge at finite $t$.
\begin{figure}[htb]
\epsfig{file=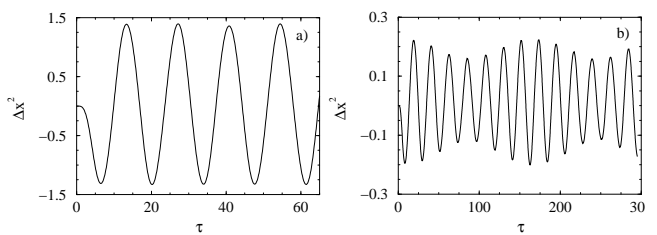,width=8.5cm}
\vspace*{-0.5cm}
\caption{Time evolution of the breathing mode $\Delta x^2$ in a trap with 
$m \omega^2 a^2 = 0.5$,
$\mu = 0.8$, a) for $t/U = 0.23$, and b) for $t/U = 0.1$. Length and time are 
measured in units of $a$ and $\hbar/U$ respectively.}
\label{fig:3}
\end{figure}
In SF phase, among the low-lying collective excitations of harmonically trapped lattice bosons,
dipole oscillation describing the center of mass motion has the
lowest energy.
According to Kohn's theorem, dipole excitation energy in
a harmonic trap is $\hbar \omega$ that is independent
of two-body interaction \cite{stringari}.
In case of BHM the dipole excitation energy
deviates from this universal value due to
the presence of the optical lattice. 
For large system size and deep in the SF phase we find that it
approaches $\sqrt{2 t m} \omega a$,
in agreement with Kohn's theorem.
In experiments, dipole mode can be excited by slightly shifting the
atomic cloud from the center of the trap  by
imposing a perturbation of the form $V_{pert} \sim \sum_{i} x_{i} \hat{n}_{i}$.

Having studied the spectra in the presence of a static trap, we now 
concentrate on the effect of time-dependent perturbations of the trap
on the dynamics of BHM by solving the full time-dependent
Gutzwiller equation. As an example, we focus on an
experimentally relevant collective mode of 1D confined system,  
the breathing mode, where the center of mass of the atomic cloud remains
fixed but its size oscillates.
For harmonically trapped dilute Bose gas in quasi 1D regime, the frequency
of this oscillations is $\sqrt{3} \omega$ \cite{menotti}.
Experimental study of this mode has become an important tool for the
investigation of the correlation properties of quasi 1D quantum
gases, particularly when 1D Bose gas enters the strongly correlated 
fermionized regime \cite{menotti}.
We study the breathing mode as a linear response to a perturbing potential
of the form  $V_{pert} \sim \sum_{i} x_{i}^2 \hat{n}_{i}$. 
Oscillation frequencies of both dipole and breathing modes 
obtained from the full dynamics agree with those obtained from linearized 
time dependent method.
In Fig.3, time evolutions of the collective coordinate $\Delta x^2 =
\sum_{i}\left[\left<x_{i}^2(\tau)\right> - \left<x_{i}^2(0)\right>\right]$
are shown for two different values of
$t/U$, under the same perturbation strength.
We observe that the response of the system becomes smaller
as it approaches the insulating regime. 
The MI phase hardly responds to perturbations due to the energy gap
in the spectrum \cite{excitation}. 
In the SF phase with small $t/U$ (close to the MI boundary), the breathing mode
oscillations show beats instead of a single frequency. As noted earlier,
in this regime excitation energies are very close to each other
and many avoided level crossings occur (see Fig. 2). Due to this reason
mode coupling takes place leading to beats in the
dynamics of the collective modes.

In conclusion, we develop a new time-dependent variational method to obtain
the spectra of correlated lattice bosons in an (arbitrary) confining
potential. In the homogeneous phase, we find that, at commensurate filling
and at the SF-MI boundary, the sound velocity does not vanish even though
the SF order parameter vanishes. Low-lying excitations of BHM in a 
harmonic trap are obtained,
which validates the GPE approach asymptotically, while
deviating significantly in the correlated regime. The dynamics of the
collective coordinates under time-dependent perturbations reveals
decrease in their amplitudes as well as appearance
of beats in the strongly correlated regime. These results 
are relevant to the experimental observation of coexisting MI-SF phases
and measurements of their dynamical properties.

The work of DLK was supported by EPSRC grant EP/D066379/1. This work was
also supported by IFRAF institute. DLK wishes to thank D. Jaksch for helpful
comments and discussions.


\end{document}